\documentclass[12pt]{article} 
\usepackage[margin=1in]{geometry}
\usepackage{hyperref}
\usepackage{fancyhdr}
\usepackage{titling}
\usepackage{authblk}
\usepackage{algorithm}
\usepackage{algorithmicx}
\usepackage{algpseudocode}
\usepackage{multirow}
\usepackage{booktabs}
\usepackage{placeins}
\usepackage{amsmath, amssymb}
\usepackage{graphicx}
\usepackage{setspace}
\usepackage{float}

\usepackage{abstract}
\newcommand{\ieeepreprintnotice}{%
  \vspace{1em}\noindent\footnotesize
  \textbf{Preprint notice.} This work has been accepted to IEEE EMBC 2025 (14–17 July 2025, Copenhagen, Denmark). Proceedings expected in December 2025.
}

\title{LWT-ARTERY-LABEL: A Lightweight Framework for Automated Coronary Artery Identification}

\author[1]{Shisheng Zhang}
\author[1]{Ramtin Gharleghi}
\author[2]{Sonit Singh}
\author[3]{Daniel Moses}
\author[4]{Dona Adikari}
\author[2]{Arcot Sowmya}
\author[1]{Susann Beier}

\affil[1]{School of Mechanical and Manufacturing Engineering, University of New South Wales, Australia}
\affil[2]{School of Computer Science and Engineering, University of New South Wales, Australia}
\affil[3]{Department of Radiology, Prince of Wales Hospital, Sydney, Australia}
\affil[4]{School of Medicine, University of New South Wales, Australia}

\date{}

\begin{document}
\maketitle

\ieeepreprintnotice

\begin{abstract}
\large
Coronary artery disease (CAD) remains the leading cause of death globally, with computed tomography coronary angiography (CTCA) serving as a key diagnostic tool. However, coronary arterial analysis using CTCA, such as identifying artery-specific features from computational modelling, is labour-intensive and time-consuming. Automated anatomical labelling of coronary arteries offers a potential solution, yet the inherent anatomical variability of coronary trees presents a significant challenge. Traditional knowledge-based labelling methods fall short in leveraging data-driven insights, while recent deep-learning approaches often demand substantial computational resources and overlook critical clinical knowledge. To address these limitations, we propose a lightweight method that integrates anatomical knowledge with rule-based topology constraints for effective coronary artery labelling. Our approach achieves state-of-the-art performance on benchmark datasets, providing a promising alternative for automated coronary artery labelling.
\end{abstract}

{\large  % or \Large depending on how big you want
\setstretch{1.5}

\section{Introduction}
Coronary Artery Disease (CAD) is the leading cause of death worldwide \cite{roth2017global} with Computed Tomography Coronary Angiography (CTCA) being the preferred diagnostic tool to assess the severity of coronary artery narrowing \cite{leipsic2014scct}. In practice, coronary arterial analysis, such as identifying artery-specific features from computational modelling, can be time-consuming. An automated anatomical labelling tool would be highly beneficial for managing large screening populations in both clinical and biomedical engineering tasks. The coronary artery tree consists of two main components: the left and right coronary arteries, both of which originate from the aorta \cite{loukas2013clinical}. The primary coronary arteries of interest include the Left Main coronary artery (LM), Left Anterior Descending artery (LAD), Left Circumflex artery (LCx), Ramus Intermedius (RI), Diagonal branches (D), Obtuse Marginal branches (OM), Septal branches (Sep), Right Coronary Artery (RCA), and Acute Marginal artery (AM), as shown in Figure \ref{4.1.F1}. The LAD, LCx, and RI originate from the LM; the Sep and D branches originate from the LAD; OM from LCx; and AM from RCA. These relationships form structured data, with LM, LAD, LCx, and RCA identified as the main arteries, while the others are side branches.

Several methods for coronary artery anatomical labelling have been developed \cite{yang2011automatic, cao2017automatic}. However, significant anatomical variability among individuals poses difficulties for labelling systems. The number of branches, the length and size of each branch, and the orientation of branches vary greatly among individuals. Previous work has relied on registration algorithms and prior knowledge \cite{yang2011automatic, cao2017automatic}. These methods typically identify four main arteries (LM, LAD, LCx, and RCA) before labelling the side branches (e.g., D, OM, AM, etc.). The results are then refined using logical rules derived from clinical experience. While effective to an extent, these conventional methods are not data-driven, and therefore cannot leverage big data to enhance accuracy. 

Deep learning is a subset of machine learning which uses deep neural networks to learn features from large datasets \cite{goodfellow2016deep}. This data-driven approach can overcome the limitations of traditional methods which often rely on manual feature extraction or rule-based algorithms that can be time-consuming, resource-intensive and less robust with large, diverse datasets. Some recent studies have used models such as 3D ResUNet with graph convolutional networks or Conditional Partial-Residual Graph Convolutional Networks for artery labelling \cite{zhang2023topology, yang2020cpr}. Although these models are data-driven, their training requires substantial computational resources and deploying such models with existing CAD software poses challenges. Furthermore, they overlook the incorporation of clinically proven prior knowledge, which could enhance their effectiveness and reliability. 

In this paper, we propose a neural network with rule-based topology constraints that leverages a set of two geometric and four spatial features from extracted centrelines. This design makes the neural network efficient and lightweight, allowing easy integration with other coronary arterial analysis and processing software. The rule-based topology constraint is applied to positional features to reduce errors. Additionally, we benchmarked our method against existing techniques, demonstrating that it achieves state-of-the-art performance.

\section{Methods}
We present our proposed method (Figure \ref{4.1.F1}), which employs a neural network to obtain labelling probabilities from geometric and spatial features and incorporates topology constraints using a rule-based model.

\begin{figure}[!htbp]
    \centering
    \includegraphics[width=0.75\linewidth]{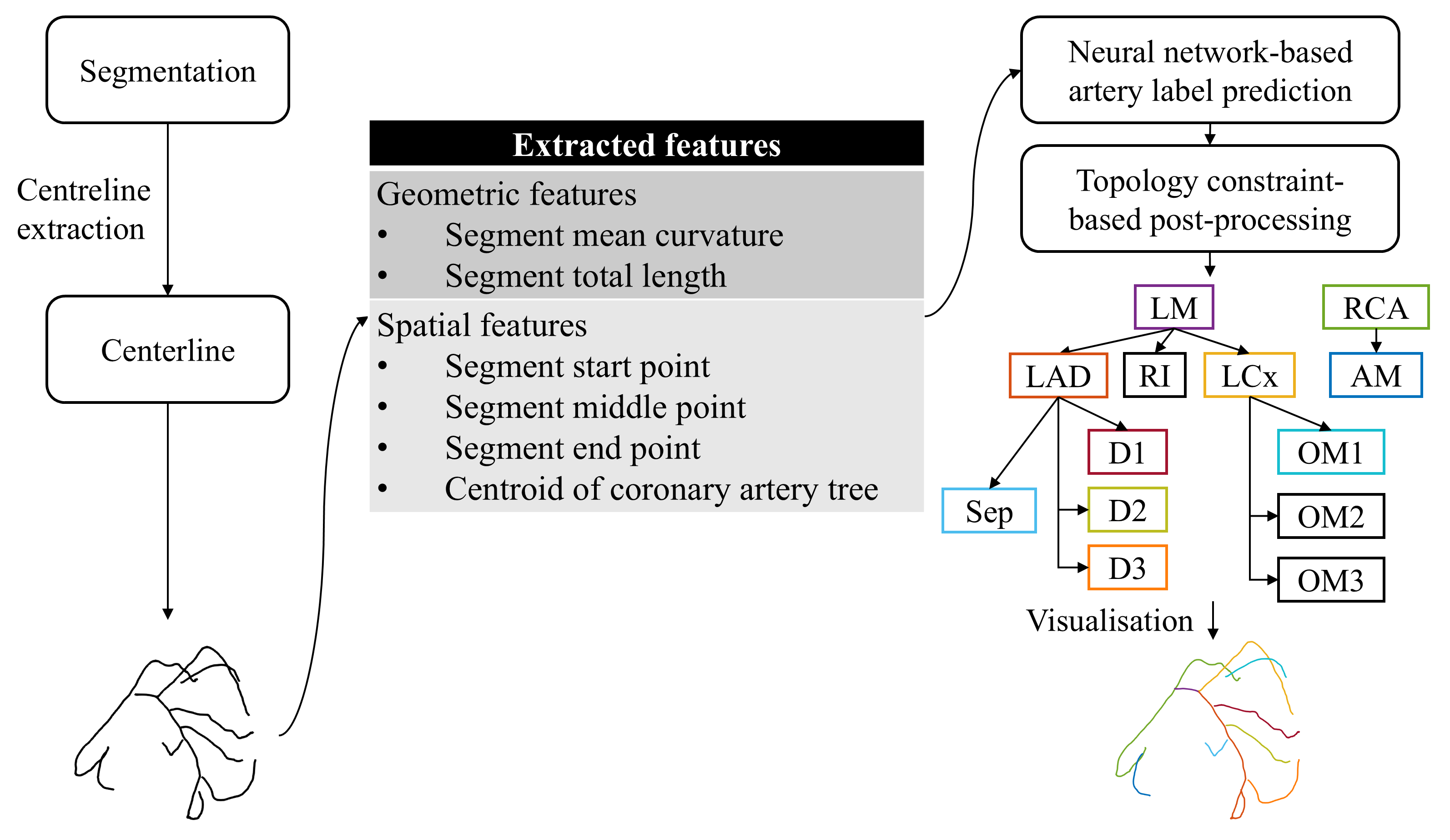}
    \caption[Flowchart of proposed method]{Flowchart of proposed method. We obtained coronary artery segmentation from CTCA and used VMTK to extract centrelines and geometric features. After that, we used neural networks to predict artery labels and applied topology constraint-based post-processing. The final output was visualised in different colours.}
    \label{4.1.F1}
\end{figure}
\FloatBarrier

\subsection{Neural network-based model for artery labelling}
Our approach to artery labelling employs a centreline skeleton model that utilises the length and curvature of arteries. These skeleton models were derived from the segmentation of medical images. To ensure the broad applicability of our method, we do not impose specific requirements on the segmentation and skeletonisation techniques used. This is also an efficient and lightweight approach, requiring fewer features and shallow neural networks to obtain the probability of an artery belonging to a candidate label (Table \ref{4.1.T1}). The network includes an input layer, four hidden layers, and an output layer. The input layer takes 14 variables and expands to 50 dimensions, followed by batch normalisation. The hidden layers adjust dimensions: 100, 75 and 50, each applying batch normalisation and ReLU activation. A dropout layer (20\%) follows to enhance generalisation. The output layer maps the features to 13 candidate labels.

Main coronary branches are common, and smaller branches may not be present in all cases. Thus, some branch labels are less frequent than major ones, resulting in an imbalanced dataset. To address this, we employ the focal loss function, which emphasises learning from hard-to-classify examples and mitigates class imbalance \cite{lin2017focal}. Focal loss modifies the standard cross-entropy loss by incorporating a modulating factor that reduces the impact of well-classified examples, focusing more on misclassified ones. We set the gamma parameter to 2 for Focal loss.

\begin{table}[h!]
\centering
\caption[Neural Network Structure for Artery Labelling]{Neural Network Structure for Artery Labelling}
\label{4.1.T1}
\resizebox{\textwidth}{!}{%
\begin{tabular}{lccl}
\hline
\textbf{Layer} & \textbf{Input dimension} & \textbf{Output dimension} & \textbf{Normalisation/Activation} \\
\hline
Input     & 14  & 50  & Batch Normalisation, ReLU \\
Hidden 1  & 50  & 100 & Batch Normalisation, ReLU \\
Hidden 2  & 100 & 75  & Batch Normalisation, ReLU \\
Hidden 3  & 75  & 50  & Batch Normalisation, ReLU \\
Dropout   & 50  & 50  & 20\% dropout \\
Output    & 50  & 13  & Softmax \\
\hline
\end{tabular}%
}
\end{table}

Coronary artery segmentations are transformed into the Right-Anterior-Superior coordinate system for consistency. We use the Vascular Modelling Toolkit (VMTK) to extract the centreline and derive geometric features from arterial segments \cite{antiga2008image}. First, we employ the vmtkcenterlines module, which processes the vascular surface to compute the centreline geometry. This involves inputting the vascular segmentation and defining inlet and outlet points to guide the computation of the central path along each vessel. Subsequently, we use the vmtkcenterlinegeometry module to analyse the extracted centreline data and derive specific geometric features, including the total length and the mean curvature for each artery segment. Spatial features calculated include the Cartesian coordinates of each artery’s start, middle, and end points and the centroid of the whole coronary artery tree, were calculated considering the CT scan spacing parameters and normalised so that the origin coordinates of the CT scan map to zero. This normalisation ensures that our method is invariant to the location of skeletons in CT volumes, eliminating the need for global alignment. Consequently, for each segment $i$, we obtain the probability

\[
P_i = [P_i^{\mathrm{LM}}, P_i^{\mathrm{LAD}}, P_i^{\mathrm{LCx}}, P_i^{\mathrm{RI}}, P_i^{\mathrm{D1}}, P_i^{\mathrm{D2}}, P_i^{\mathrm{D3}}, 
        P_i^{\mathrm{OM1}}, P_i^{\mathrm{OM2}}, P_i^{\mathrm{OM3}}, P_i^{\mathrm{Sep}}, P_i^{\mathrm{RCA}}, P_i^{\mathrm{AM}}]
\]

where $P_i$ is the probability vector that represents the probability of the segment $i$ belonging to a specific artery category. 

\subsection{Rule-based topological constraints for post-processing}
Human vascular anatomy, including the coronary arteries, typically exhibits a specific order of vessel arrangement \cite{loukas2013clinical}. For instance, the common diagonal (D1/D2/D3) and ostium marginal (OM1/OM2/OM3) arteries are generally located downstream from the LAD and LCx arteries, pointing downward to supply blood to the heart muscle. Despite the significant variability in morphology and topology, imposing such prior knowledge as topological constraints can be valuable during the inference process. Thus, we formulated the labelling task as a probability model, whereby the post-processing involves constructing a major artery graph based on prior probabilities using probability values for each branch.

The first step is to distinguish between the left and right coronary arterial trees. Typically, in the Right-Anterior-Superior coordinate system, the left coronary artery is located on the positive side of the x-axis, while the right coronary artery is on the negative side. First, we calculated the centroid of all arteries (i.e. right and left coronary arteries and all of their branches) and the midpoint of each artery segment. If the x-coordinate of a segment’s midpoint is less than the centroid’s x-coordinate, the segment belongs to the right coronary arterial tree; otherwise, it belongs to the left coronary arterial tree.  This is implemented in Lines 1-4 in Algorithm 1. 

The second step is to identify the RCA and AM arteries based on prior probability. If multiple RCA segments are present, the one with the highest probability is selected, and the same method applies to AM. This is in Lines 5-8. 

In the third step, we assign the labels for the LM, LAD, and LCx arteries. Each label is assigned to the segment with the highest corresponding probability. This is implemented in Lines 9-13.

The RI artery is typically found in the proximal area between LAD and LCx in clinical studies \cite{kini2007normal, fioranelli2013clinical}. The fourth step in our process involves verifying the prediction of the RI’s presence. If RI is predicted, we then assess whether the distance from the start of RI to the end of LM is below a predefined threshold, which we have set at 3 mm. If this distance falls within the threshold, the segment is confirmed as RI and labelled accordingly. If the distance exceeds the threshold, we then compare the probabilities of the segment being D1 or OM1. The segment’s initial RI prediction is subsequently replaced with D1 or OM1, depending on which has the higher probability. This is implemented in Lines 14-20. In the fifth step, Diagonal branches (D1, D2, D3) are identified by their connection order to the LAD, correcting any discrepancies (Lines 21-23). Similarly, in the sixth step, Ostium Marginal branches (OM1, OM2, OM3) are identified by their connection to the LCx, with corrections made if needed (Lines 24-26).

Finally, in the seventh step, the method checks if the Septal artery (Sep) was predicted. If so, the Sep label is assigned accordingly. This is implemented in Lines 27-29. 

\newpage
\vspace*{-3em}
\begin{algorithm}[H]
\caption{Rule-based Topological Constraints for Post-Processing}
\label{alg:topo_constraints}
\textbf{Input:} set of segment index $S = \{1, \ldots, N\}$, set of predicted labels $\hat{Y} = \{\hat{y}_1, \ldots, \hat{y}_N\}$, where $N$ is the total number of segments. For each segment $i \in S$, probability vector across all artery categories $P_i = [P_i^{LM}, P_i^{LAD}, P_i^{LCx}, P_i^{RI}, P_i^{D1}, P_i^{D2}, P_i^{D3}, P_i^{OM1}, P_i^{OM2}, P_i^{OM3}, P_i^{Sep}, P_i^{RCA}, P_i^{AM}]$ \\
\textbf{Output:} Artery labels for each segment $LM$, $LAD$, $LCx$, $RI$, $D1$, $D2$, $D3$, $OM1$, $OM2$, $OM3$, $Sep$, $RCA$, $AM$
{\setstretch{0.7}
\begin{algorithmic}[1]

\State \textbf{Step 1: Create right and left set}
\State let $x_{i,mid}$ = x-coordinate of the midpoint of segment $S_i$
\State let $x_c$ = x-coordinate of the centroid of the coronary artery tree
\State divide segments into right set $S_{Right} = \{i \in S \mid x_{i,mid} < x_c\}$ and left set $S_{Left} = \{i \in S \mid x_{i,mid} \geq x_c\}$

\State \textbf{Step 2: Assign RCA and AM from right set}
\State find $RCA$ and $AM$ predicted segments from right set $S_{RCA} = \{i \in S_{Right} \mid \hat{y}_i = RCA\}$ and $S_{AM} = \{i \in S_{Right} \mid \hat{y}_i = AM\}$, respectively
\State $RCA = 
    \begin{cases}
        S_{\text{RCA}} & \text{if } \lvert S_{\text{RCA}} \rvert = 1 \\
        \arg\max\limits_{i \in S_{\text{RCA}}} P_i^{\text{RCA}} & \text{if } \lvert S_{\text{RCA}} \rvert > 1
    \end{cases}$
\State $AM = 
    \begin{cases}
        \emptyset & \text{if } \lvert S_{\text{AM}} \rvert = 0 \\
        S_{\text{AM}} & \text{if } \lvert S_{\text{AM}} \rvert = 1 \\
        \arg\max\limits_{i \in S_{\text{AM}}} P_i^{\text{AM}} & \text{if } \lvert S_{\text{AM}} \rvert > 1 
    \end{cases}$

\State \textbf{Step 3: Assign LM, LAD and LCx from left set}
\State find $LM$, $LAD$ and $LCx$ predicted segments from left set $S_{LM} = \{i \in S_{Left} \mid \hat{y}_i = LM\}$, $S_{LAD} = \{i \in S_{Left} \mid \hat{y}_i = LAD\}$ and $S_{LCx} = \{i \in S_{Left} \mid \hat{y}_i = LCx\}$, respectively
\State $LM = 
    \begin{cases}
        S_{\text{LM}} & \text{if } \lvert S_{\text{LM}} \rvert = 1 \\
        \arg\max\limits_{i \in S_{\text{LM}}} P_i^{\text{LM}} & \text{if } \lvert S_{\text{LM}} \rvert > 1
    \end{cases}$
\State $LAD = 
    \begin{cases}
        S_{\text{LAD}} & \text{if } \lvert S_{\text{LAD}} \rvert = 1 \\
        \arg\max\limits_{i \in S_{\text{LAD}}} P_i^{\text{LAD}} & \text{if } \lvert S_{\text{LAD}} \rvert > 1
    \end{cases}$
\State $LCx = 
    \begin{cases}
        S_{\text{LCx}} & \text{if } \lvert S_{\text{LCx}} \rvert = 1 \\
        \arg\max\limits_{i \in S_{\text{LCx}}} P_i^{\text{LCx}} & \text{if } \lvert S_{\text{LCx}} \rvert > 1
    \end{cases}$

\State \textbf{Step 4: Assign RI from left set}
\State $S_{RI} = \{i \in S_{Left} \mid \hat{y}_i = \text{RI}\}$
\State let $RI_{\text{start}}$ = coordinate of the start point of segments in $S_{RI}$
\State let $LM_{\text{end}}$ = coordinate of the end point of $LM$
\State let $RI_{\text{threshold}}$ be a clinically defined maximum distance between LM end point and RI start point
\State $S_{RI} = \{i \in S_{Left} \mid \hat{y}_i = \text{RI} \text{ and } \text{dist}(RI_{\text{start}}, LM_{\text{end}}) < RI_{\text{threshold}}\}$
\State $RI = 
    \begin{cases}
        \emptyset & \text{if } \lvert S_{\text{RI}} \rvert = 0 \\
        S_{\text{RI}} & \text{if } \lvert S_{\text{RI}} \rvert = 1 \\
        \arg\max\limits_{i \in S_{\text{RI}}} P_i^{\text{RI}} & \text{if } \lvert S_{\text{RI}} \rvert > 1 
    \end{cases}$

\State \textbf{Step 5: Assign Diagonal branches from left set}
\State $S_{D} = \{i \in S_{Left} \mid \hat{y}_i = D_1 \text{ or } D_2 \text{ or } D_3\}$
\State $\{D_1, D_2, D_3\} = \text{sort}(S_D, \text{key} = \text{position\_on\_LAD})$

\State \textbf{Step 6: Assign Ostium Marginal branches from left set}
\State $S_{OM} = \{i \in S_{Left} \mid \hat{y}_i = OM_1 \text{ or } OM_2 \text{ or } OM_3\}$
\State $\{OM_1, OM_2, OM_3\} = \text{sort}(S_{OM}, \text{key} = \text{position\_on\_LCx})$

\State \textbf{Step 7: Assign Septal artery from left set}
\State $S_{Sep} = \{i \in S_{Left} \mid \hat{y}_i = Sep\}$
\State $Sep = 
    \begin{cases}
        \emptyset & \text{if } \lvert S_{\text{Sep}} \rvert = 0 \\
        S_{\text{Sep}} & \text{if } \lvert S_{\text{Sep}} \rvert = 1 \\
        \arg\max\limits_{i \in S_{\text{Sep}}} P_i^{\text{Sep}} & \text{if } \lvert S_{\text{Sep}} \rvert > 1 
    \end{cases}$
\end{algorithmic}
}
\end{algorithm}
\thispagestyle{empty}

\subsection{Datasets}
Our study utilised three CTCA datasets. The first dataset, the Coronary Atlas, includes 380 annotated cases sourced from Intra Imaging, Auckland, New Zealand \cite{medrano2014construction, medrano2016computational}. The study was approved by the institutional review committee and all subjects provided written informed consent. The second dataset is an internal test set comprising 70 patients with suspected CAD from the GeoCAD dataset \cite{adikari2022new}. Approval was granted by the St Vincent’s Hospital Human Research Ethics Committee, Sydney (Ref. 2020/ETH02127) and the NSW Population and Health Service Research Ethics Committee (Ref. 2021/ETH00990). Finally, an external dataset, namely orCaScore \cite{wolterink2016evaluation}, comprising 72 annotated cases was also employed for testing purposes \cite{zhang2023topology}. This study was performed in line with the principles of the Declaration of Helsinki. 

\subsection{Implementation and Evaluation}\label{Implementation and Evaluation}
For the implementation, we used Python 3.10, PyTorch 2.0 \cite{paszke2017automatic}, Scikit-Learn 1.2 \cite{pedregosa2011scikit}, and Numpy 1.24 \cite{harris2020array} on Windows 11 to develop the neural networks and integrate topology constraints. The CPU and GPU used are i7 13700KF and NVIDIA GeForce RTX 3090 Ti, respectively. The neural networks were trained for 3,000 epochs with a batch size of 64. The evaluation metrics used to assess the method’s performance included recall, precision, and F1-score \cite{bishop2006pattern}. We employed five-fold cross-validation to train our model on the Coronary Atlas dataset. The evaluation was conducted using the internal dataset GeoCAD and the external dataset orCaScore for benchmarking purposes, with mean probabilities from five trained folds refined by applying rule-based topological constraints. The performance of our proposed method was evaluated by comparing it against several state-of-the-art approaches, including TreeLab-Net \cite{wu2019automated}, CPR-GCN \cite{yang2020cpr}, CorLab-Net \cite{zhang2021corlab}, TaG-Net \cite{yao2023tag}, and TopoLab \cite{zhang2023topology}.

\section{Results}
Our method performed well on both the internal and external test sets, as summarised in Table \ref{4.1.T2}. On the internal test set, GeoCAD, our model achieved an average recall of 93.06\%, precision of 93.41\%, and F1-score of 93.17\%. On the external test set, orCaScore, the results were similarly robust, with a recall of 93.21\%, precision of 94.00\%, and F1-score of 93.26\%. Most arteries, including the LM, LAD, LCx, RI, D1, OM1, Sep, RCA and AM, demonstrated recall, precision and F1 score around 90\%. However, these metrics were relatively lower for smaller branches, such as the second and third diagonal branches (D2 and D3) and the second and third obtuse marginal branches (OM2 and OM3). 

\begin{table}[H]
\centering
\caption{Evaluation of our method on GeoCAD and orCaScore datasets (\%). AM was not included in GeoCAD labelling. Each row shows the recall, precision and F1 score}
\label{4.1.T2}
\resizebox{\textwidth}{!}{%
\begin{tabular}{ccccccc}
\hline
\multirow{2}{*}{} & \multicolumn{3}{c}{\textbf{GeoCAD}} & \multicolumn{3}{c}{\textbf{orCaScore}} \\
\cmidrule(lr){2-4} \cmidrule(lr){5-7}
 & \textbf{Recall} & \textbf{Precision} & \textbf{F1 score} & \textbf{Recall} & \textbf{Precision} & \textbf{F1 score} \\
\hline
LM   & 95.45 & 98.44 & 96.92 & 100.0 & 100.0 & 100.0 \\
LAD  & 96.97 & 92.75 & 94.81 & 100.0 & 100.0 & 100.0 \\
LCx  & 93.94 & 89.86 & 91.85 & 100.0 & 100.0 & 100.0 \\
RI   & 94.44 & 94.44 & 94.44 & 96.00 & 60.00 & 73.85 \\
D1   & 92.98 & 94.64 & 93.81 & 94.20 & 92.86 & 93.53 \\
D2   & 76.47 & 81.25 & 78.79 & 83.33 & 88.89 & 86.02 \\
D3   & 66.67 & 66.67 & 66.67 & 72.22 & 92.86 & 81.25 \\
OM1  & 86.36 & 95.00 & 90.48 & 84.38 & 88.52 & 86.40 \\
OM2  & 81.82 & 90.00 & 85.71 & 73.53 & 89.29 & 80.65 \\
OM3  & 100.0 & 100.0 & 100.0 & 57.14 & 100.0 & 72.73 \\
Sep  & 94.12 & 80.00 & 86.49 & 97.30 & 90.00 & 93.51 \\
RCA  & 96.77 & 98.36 & 97.56 & 100.0 & 100.0 & 100.0 \\
AM   & -     & -     & -     & 100.0 & 100.0 & 100.0 \\
\hline
\textbf{Avg.} & \textbf{93.06} & \textbf{93.41} & \textbf{93.17} & \textbf{93.21} & \textbf{94.00} & \textbf{93.26} \\
\hline
\end{tabular}%
}
\end{table}

\begin{figure}[!htbp]
    \centering
    \includegraphics[width=1.0\linewidth]{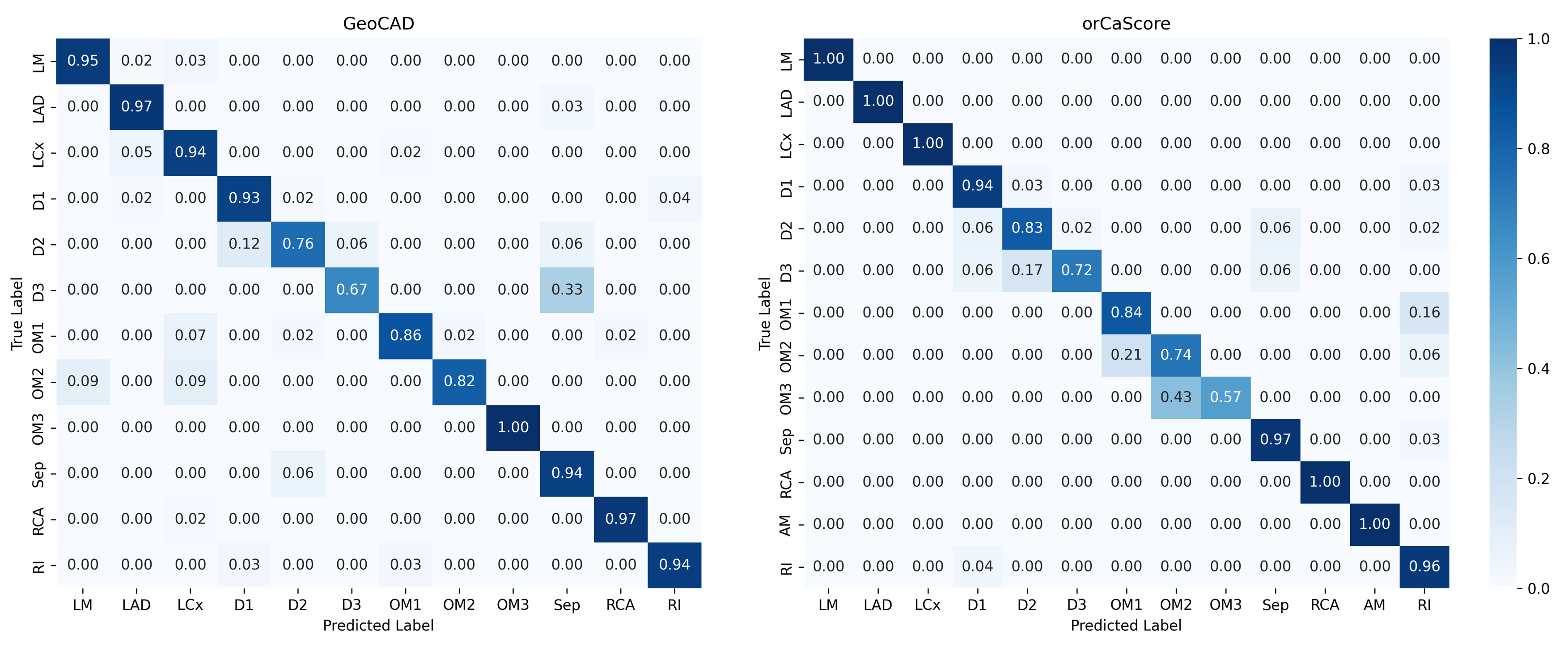}
    \caption[Confusion matrices for GeoCAD and orCaScore labelling]{Confusion matrices for GeoCAD (left) and orCaScore (right) labelling. Each row represents the percentage of predictions, with the diagonal numbers indicating the percentage of correct predictions.}
    \label{4.1.F2}
\end{figure}
\FloatBarrier

The confusion matrices (Figure \ref{4.1.F2}) for 13 classes of branches display the distributions of the resulting labels. For the GeoCAD dataset, two noticeable missing branches were D2 and D3, with 12\% of D2 and 33\% of D3 incorrectly labelled as D1 and Sep respectively. For the orCaScore dataset, the branches with more incorrectly assigned labels are highly related to those with fewer samples (e.g., 28\% of D3, 26\% of OM2, and 43\% of OM3 are missing). Additionally, similarly located branches (e.g., OM1 and RI) are sometimes incorrectly labelled. 

Compared to the current leading labelling method TopoLab \cite{zhang2023topology}, our method improves mean recall, precision and F1 score by 6.08\%, 5.69\%, and 6.03\% respectively on the orCaScore dataset (Table \ref{4.1.T3}). The mean metrics are weighted averages based on the number of segments in each artery category, following the approach used by Zhang et al. \cite{zhang2023topology}. This ensures that the metrics accurately reflect the distribution of segments across categories. Our method is efficient in both model size and execution time. It is a compact neural network of 72.02 KB with 18,438 parameters, which is resource-efficient while maintaining high performance using 32-bit precision. During training, the execution time for 3,000 epochs totals 263 seconds, averaging 87.98 milliseconds per epoch (experimental setup in Section \ref{Implementation and Evaluation}).

\begin{table}[H]
\centering
\caption{Comparison to other labelling methods applied to the orCaScore dataset for weighted averages of different artery categories}
\label{4.1.T3}
\resizebox{\textwidth}{!}{%
\begin{tabular}{lcccc}
\hline
\textbf{Method} & \textbf{Number of categories} & \textbf{Recall (\%)} & \textbf{Precision (\%)} & \textbf{F1 score (\%)} \\
\hline
TreeLab-Net \cite{wu2019automated}        & 10 & 83.35 & 84.90 & 83.12 \\
CPR-GCN \cite{yang2020cpr}                & 15 & 82.88 & 83.61 & 82.72 \\
CorLab-Net \cite{zhang2021corlab}          & 11 & 82.09 & 83.83 & 82.15 \\
TaG-Net \cite{yao2023tag}                & 14 & 82.29 & 83.41 & 82.14 \\
TopoLab \cite{zhang2023topology}              & 14 & 87.13 & 88.31 & 87.23 \\
\textbf{LWT-ARTERY-LABEL (Ours)}     & 13 & \textbf{93.21} & \textbf{94.00} & \textbf{93.26} \\
\hline
\end{tabular}%
}
\end{table}

\section{Discussion and Conclusion}
In this paper, we propose a neural network enhanced with rule-based topology constraints that uses geometric and spatial information from CTCA volumes to accurately predict coronary artery labels. Our method uses a small set of easily obtainable features, reducing computational power requirements. This approach can improve efficiency in the processing of medical coronary artery images towards automation of advanced computational analysis.

Our method has several limitations. First, its performance is highly dependent on the quality of the centreline generation process. Inaccuracies in centreline extraction can lead to erroneous artery labelling, affecting the overall reliability of the method. Second, to normalise coordinates, we used the CT scan origin. However, this may not align with the cardiac centre, potentially causing discrepancies if the scan includes non-cardiac regions such as the abdomen. Third, the method may not perform well with abnormal coronary artery topologies that deviate from common clinical presentations. Such cases require additional consideration to ensure accurate predictions and avoid mislabelling. 

In conclusion, our proposed approach offers a promising solution for automating the analysis of coronary arteries with minimal computational demands. Future work will focus on enhancing centreline extraction process and expanding the model’s ability to handle atypical coronary artery structures. 

}

\clearpage
\bibliographystyle{IEEEtran}
\bibliography{refs} % Remember to include the .bbl file for arXiv.
\end{document}